\documentclass[preprint,showpacs,aps,amsmath,amssymb]{revtex4}

\usepackage{bm}

\begin{document}

\preprint{Submitted to Journal of Mathematical Physics}

\title{Nonequilibrium Processes in Non-Hamiltonian Statistical Ensembles}

\author{Alexander V. Zhukov} \email{avzhukov@mit.edu}
\author{Jianshu Cao}

 \affiliation{Department of Chemistry, Massachusetts Institute of Technology,
              Cambridge, MA 02139, USA}

\date{November 10, 2003}

\begin{abstract}
A nonequilibrium statistical operator method is developed for
ensembles of particles obeying non-Hamiltonian equations of motion
in classical phase space. The main consequences of non-zero
compressibility of phase space are examined in terms of
time-dependent dynamic quantities. The generalized transport
equations involve the phase-space compressibility in a non-trivial
way. Our results are useful in molecular dynamics simulation
studies as well as  nonequilibrium or quasiclassical
approximations of quantum-classical dynamics.
\end{abstract}


\maketitle

Theoretical interest in the non-Hamiltonian motion appears in the
early works \cite{1,2,3}, where it was realized that the
constant-temperature simulations require the introduction of
additional non-Hamiltonian acceleration into dynamical equations.
This has opened new research directions associated with extended
systems. For the past few years the physics of non-Hamiltonian
systems was partially developed for molecular dynamics simulation
purposes \cite{4,5,6,cao1,cao2}. Particular applications of
non-Hamiltonian dynamical systems are overviewed in a recent paper
\cite{8}, where the authors exploit their formulation of
non-Hamiltonian statistical mechanics \cite{9}. In this regard it
is important to mention also the interesting studies of mixed
quantum-classical systems \cite{tully1,tully2,10,11,12}, where the
non-Hamiltonian behavior plays an essential role.
\par
One of the most popular non-Hamiltonian ensembles is the so-called
Nos{\'e}-Hoover chain \cite{6}. In case with two thermostat
variables, the corresponding Hamiltonian is
\begin{equation}
{\cal H}_{NH} = \frac{p^2}{2m} +\frac{p_{\eta_1}^2}{2M_{\eta_1}}
+\frac{p_{\eta_2}^2}{2M_{\eta_2}}+V(q)+gk_bT(\eta_1+\eta_2),\label{ad1}
\end{equation}
where $\eta_i$ and $p_{\eta_i}$ are the conjugate thermostat
variables, $g$ is the number of degrees of freedom, $T$ is the
thermostat temperature. It can be shown that the Hamiltonian
(\ref{ad1}) is conserved by the following (non-Hamiltonian)
equations of motion:
\begin{equation}
{\dot q}=\frac{p}{m}, \label{ad2}
\end{equation}
\begin{equation}
{\dot \eta_i}=\frac{p_{\eta_i}}{M_{\eta_i}}, \label{ad3}
\end{equation}
\begin{equation}
{\dot p}=-\frac{p_{\eta_1}}{M_{\eta_1}}p-\frac{\partial
V(q)}{\partial q}, \label{ad4}
\end{equation}
\begin{equation}
{\dot p_{\eta_1}}=\frac{p^2}{m}-gk_bT -
p_{\eta_1}\frac{p_{\eta_2}}{M_{\eta_2}}, \label{ad5}
\end{equation}
\begin{equation}
{\dot p_{\eta_2}}=\frac{p_{\eta_1}^2}{M_{\eta_1}}-gk_bT
\label{ad6}.
\end{equation}
In real MD simulations the particular choice of non-Hamiltonian
ensemble can vary depending on the physical problem under study.
But all of them can be described in general formalism we present
here. Recently using a generalization of the symplectic form of
the Hamilton equations of motion the authors of Ref.
\onlinecite{13} showed that there is a unique general structure
that underlies the majority of the equations of motion for
extended systems. Furthermore, Sergi \cite{14,ser} proposed a
consistent theory of statistical mechanics for non-Hamiltonian
systems under equilibrium conditions. The tempting feature of the
theory \cite{13,14,ser} is that it is based on the introduction of
a suitable Poisson bracket, and has the general structure of
classical formalism \cite{15} with a natural appearance of the
nonzero phase-space compressibility. The main feature of the
systems of interest is energy conservation, which is a reason of
non-Hamiltonian behavior. It is interesting therefore to consider
the situation, when the energy is still conserved, but the system
itself is out of equilibrium, particularly when the dynamical
quantities depend on time explicitly.
\par
The starting point in describing the non-Hamiltonian statistical ensembles
is the introduction of the proper non-Hamiltonian Poisson bracket:
\begin{equation}
\{A(t),C(t)\} =\sum_{ik}^{2N}
\frac{\partial A}{\partial x_i}{\cal B}_{ik}
\frac{\partial C}{\partial x_k}
\label{1},
\end{equation}
where ${\bf x}=(q,p)=(q_1,q_2...q_N;p_1,p_2...p_N)$ is the
$2N$-dimensional vector in phase space, $A(t)$ and $C(t)$ are
arbitrary dynamical quantities (we allow them to depend on time
explicitly), ${\cal B}_{ik}= -{\cal B}_{ki}$ is an antisymmetric
matrix, whose elements are functions of ${\bf x}$ and, possibly,
time. Following Ref. \onlinecite{14} we write the equations of
motion in standard form ${\dot x}_j=\{x_j,{\cal H}\}$, where
${\cal H}$ is the Hamiltonian, which is a constant of motion by
construction due to of antisymmetry of the matrix ${\cal B}_{ik}$.
\par
The main feature of the non-Hamiltonian system is the failure of
the Jacobi relation, which implies time-irreversibility of
non-Hamiltonian mechanics. This is formally a consequence of our
requirement for eq. (\ref{1}) and the standard equation of motion
to be satisfied simultaneously. By defining the Liouville operator
$$
i{\hat L}A ={\cal B}_{ij}
\frac{\partial A}{\partial x_i}
\frac{\partial {\cal H}}{\partial x_j}=
\{A,{\cal H}\},
$$
we obtain the generalized Liouville equation (see Ref. \onlinecite{13}) in the form
\begin{equation}
\frac{\partial}{\partial t}\rho (q,p;t)= - \left(i{\hat L} +
\varkappa \right)\rho (q,p;t)
\label{3},
\end{equation}
where
$$
\varkappa =\sum_{ij}^{2N}
\frac{\partial {\cal B}_{ij}}{\partial x_i}
\frac{\partial {\cal H}}{\partial x_j}
$$
is the phase space compressibility. For the purpose of generality
we also allow it to depend on time explicitly (through ${\cal
B}_{ij}$). For particular determination of the matrix ${\cal
B}_{ij}$ (and therefore $\varkappa$) it is useful to write the
equation of motion in the form
$$
{\dot x}_i = \sum_{j=1}^{2N}{\cal B}_{ij}\frac{\partial {\cal
H}}{\partial x_j}.
$$
For the particular ensemble, described by Eqs. (\ref{ad1}) -
(\ref{ad6}), we obtain
$$
{\cal B} = \left( \begin{array}{cccccc}
0 & 0 & 0 & 1 & 0 & 0 \\
0 & 0 & 0 & 0 & 1 & 0 \\
0 & 0 & 0 & 0 & 0 & 1 \\
-1 & 0 & 0 & 0 & -p & 0 \\
0 & -1 & 0 & p & 0 & -p_{\eta_1} \\
0 & 0 & -1 & 0 & p_{\eta_1} & 0
\end{array}\right),
$$
so that the compressibility can be calculated as
$$
\varkappa = -\frac{p_{\eta_1}}{M_{\eta_1}}-
\frac{p_{\eta_2}}{M_{\eta_2}}.
$$

\par
In general for a system of $N$ identical particles (more exactly
for $N$ degrees of freedom) we define the statistical
time-dependent average as
\begin{equation}
\langle A\rangle^t = \sum_N\int A_N(q,p;t)\rho_N(q,p;t)d\Gamma_N
\equiv {\rm Tr}\{ A(q,p;t)\rho(q,p;t)\}
\label{5},
\end{equation}
where $A_N(q,p;t)$ denotes the dynamic variable $A$ for a system of $N$ particles,
i.e., $A(q_1,...,q_N;p_1,...,p_N;t)$, $d\Gamma_N$ is the element of phase volume.
\par
The time evolution of an arbitrary dynamical variable obeys the usual equation
\begin{equation}
\frac{dA}{dt}=\frac{\partial A}{\partial t}+ \{A,{\cal H}\}
=\frac{\partial A}{\partial t}+i{\hat L}A
\label{6}.
\end{equation}
One interesting consequence of the time-irreversibility is that the average derivative
of a dynamical variable is not necessarily equal to the derivative of its average.
In fact, taking into account eq. (\ref{3}) we obtain the relation
\begin{equation}
\frac{d\langle A\rangle^t}{dt}=\int
\left( \frac{\partial A}{\partial t} +\{A,{\cal H}\} - \varkappa A
\right)\rho d\Gamma
\label{7},
\end{equation}
or, alternatively
\begin{equation}
\Biggl\langle\frac{dA}{dt}\Biggr\rangle^t=
\frac{d\langle A\rangle^t}{\partial t}+
\varkappa\langle A\rangle^t
\label{8}.
\end{equation}
Equations (\ref{7}) and (\ref{8}) are particularly important in
identifying the real measured mean quantity. Another important
peculiarity of non-Hamiltonian systems follows if one is
interested in the time evolution operator for the initial
statistical distribution $\rho
(q,p;t)=U_{\pm}^{(\varkappa)}(t,t')\rho(q,p;0)$. Following the
standard procedure \cite{16} we obtain the time-ordered evolution
operator $U_{\pm}^{(\varkappa)}(t,t')$ in the following form
\begin{equation}
U_{\pm}^{(\varkappa)}(t,t')=\exp_{\pm}\left\{
-\int\limits_{t'}^t\left( i{\hat L}(\tau)+\varkappa\right)d\tau
\right\}
\label{9}
\end{equation}
where $\exp_{\pm}$ denote the correspondingly time-ordered exponents.
We emphasize that the evolution operator in Eq. (\ref{9}) does not describe
the evolution of dynamic variables. Eq. (\ref{6}) gives
$A(q,p;t)=U_{\pm}^\dag(t,t')A(q,p;t')$, where
\begin{equation}
U_{\pm}^\dag(t,t')=U_{\pm}^{(0)\dag}(t,t')
=\exp_{\pm}\left\{ i\int\limits_{t'}^t{\hat L}(\tau)d\tau
\right\}
\label{10}.
\end{equation}
\par
Let us suppose that the given nonequilibrium non-Hamiltonian ensemble
can be characterized by the average values $\langle \zeta_n \rangle^t$ of
a finite set of dynamical variables $\{\zeta_n(t)\}$. Then the time evolution
of the nonequilibrium macroscopic state is described by the generalized
transport equations (see Eqs. (\ref{7}) and (\ref{8}))
\begin{equation}
\frac{\partial\langle \zeta_n \rangle^t}{\partial t}=
{\rm Tr}\left\{\left(\dot\zeta_n - \varkappa\zeta_n\right)\rho(t)\right\}
={\rm Tr}\left\{\left(\{\zeta_n,{\cal H}\} - \varkappa\zeta_n\right)\rho(t)\right\}
\label{11},
\end{equation}
where the trace operation is defined in Eq. (\ref{5}). Let $\tilde\rho(t)$ be a trial
statistical operator, such that
$\langle \zeta_n \rangle^t={\rm Tr}\left(
\zeta_n\tilde\rho(t)
\right)$, and $ {\rm Tr}\left(
\tilde\rho(t)\right)=1$.
We choose the entropy functional in the standard form \cite{16},
\begin{equation}
\tilde S [\tilde\rho(t)]= -{\rm Tr}\left(
\tilde\rho(t)\ln \tilde\rho(t)\right)
-\sum_n\varphi_n(t) {\rm Tr} \left(\zeta_n\tilde\rho(t)\right)
- \lambda(t){\rm Tr}\tilde\rho(t)
\label{12},
\end{equation}
where $\varphi_n(t)$ and $\lambda(t)$ are the corresponding Lagrange multipliers.
Varying (\ref{12}) with
respect to the trial distribution $\tilde\rho(t)$ we obtain the {\it relevant}
distribution,
\begin{equation}
\rho_r=\exp\left\{ - \Phi (t)-
\sum_n\varphi_n(t) \zeta_n\right\}
\label{13},
\end{equation}
where $\Phi(t)$ is the Massieu-Plank function. Here we use the
notion ``relevant'' following Zubarev \cite{16,17,18} for the
distribution that maximizes the entropy functional (\ref{12}) at a
particular time. The relevant distribution is not yet the required
nonequilibrium distribution because, in general, it does not
satisfy the Liouville equation. Nevertheless, as shown below the
relevant statistical distribution will serve as an auxiliary
distribution to determine special solutions of the Liouville
equation that describe irreversible macroscopic processes.
\par
The Lagrange multipliers $\varphi_n$ are determined from the self-consistency
conditions,
\begin{equation}
\langle\zeta_n\rangle^t = \langle\zeta_n\rangle_r^t
\equiv {\rm Tr}\left(
\zeta_n\rho_r(t) \right)
\label{14},
\end{equation}
so that $S(t) =\Phi(t)+\sum_n \varphi_n(t) \langle\zeta_n\rangle^t$.
The meaning of the nonequilibrium entropy is the same
as the corresponding entropy in the Hamiltonian systems \cite{16,17}. In contrast with the
Gibbs entropy, $S(t)$ can change in time. Let us demonstrate the
specific feature of this entropy in the non-Hamiltonian system by
calculating the derivative
\begin{equation}
\frac{dS(t)}{dt}=\sum_n\frac{\delta S(t)}{\delta\langle\zeta_n\rangle^t}
\frac{\partial\langle\zeta_n\rangle^t}{\partial t}=
\sum_n\varphi_n
\frac{\partial\langle\zeta_n\rangle^t}{\partial t}
=\sum_n\varphi_n {\rm Tr}\left\{
\left( \dot\zeta_n - \varkappa\zeta_n \right)\rho(t)
\right\}.
\label{15}
\end{equation}
It seems confusing that $dS(t)/dt\neq 0$ even for constant
$\zeta_n$ due to the compressibility of the phase space. This of
course is misunderstanding, because the entropy and its evolution
are essentially determined by the {\it mean values} of $\zeta_n$
and by the derivatives of their mean values. This is particularly
important in applications of the theory since the interpretations
of the desired quantities depend on the definition of the
averaging procedure.
\par
To find the formal solution of the generalized Liouville equation (\ref{3}),
we assume that at some initial instant of time
the distribution coincides with the relevant one (\ref{13}), satisfying
the self-consistency relations (\ref{14}). Following the method proposed by
Zubarev \cite{18}, the retarded solutions of the Liouville equation in our
case is found by solving the following {\it perturbed} equation
\begin{equation}
\frac{\partial \rho(t)}{\partial t} + \left(
i{\hat L}+\varkappa \right)\rho (t) =
-\epsilon \left( \rho(t) - \rho_r(t)\right)
\label{16}
\end{equation}
with $\epsilon\rightarrow 0$ in the final expressions. The formal
solution of the initial Liouville equation (\ref{3}) reads
\begin{equation}
\rho(t)=\lim_{\epsilon\rightarrow +0}\epsilon\int\limits_{-\infty}^t
e^{-(\epsilon+\varkappa)(t-t')}e^{-i(t-t'){\hat L}}\rho_r(t')dt'
\label{17}.
\end{equation}
Integrating eq. (\ref{17}) by parts we arrive at
\begin{equation}
\rho(t)=\rho_r(t)-
\lim_{\epsilon\rightarrow +0}\epsilon\int\limits_{-\infty}^t
e^{-(\epsilon+\varkappa)(t-t')}e^{-i(t-t'){\hat L}}
\left(\frac{\partial}{\partial t'}+i{\hat L}+\varkappa\right)\rho_r(t')dt'
\label{18}.
\end{equation}
In principle equation (\ref{18}) answers the question how to construct nonequilibrium
distribution, which is a functional of dynamic variables and the phase space
compressibility, if the system develops from a relevant state.
\par
Let us further write the statistical operator as a sum of relevant
distribution and local-time deviation
$\rho(t)=\rho_r(t)+\Delta\rho(t)$, so that equation (\ref{16})
becomes
\begin{equation}
\left(
\frac{\partial}{\partial t}+i{\hat L} + \varkappa +\epsilon
\right)\Delta\rho(t)=-
\left(
\frac{\partial}{\partial t}+i{\hat L} + \varkappa \right)
\rho_r(t)
\label{19}.
\end{equation}
Then, keeping in mind that $\rho_r(t)$ depends on time only through
the relevant variables $\zeta_n$, we can calculate its time derivative
\begin{eqnarray}
\frac{\partial\rho_r(t)}{\partial t}=
\sum_n\frac{\delta\rho_r(t)}{\delta\langle\zeta_n\rangle^t}
\frac{\partial\langle\zeta_n\rangle^t}{\partial t}
=\sum_n\frac{\delta\rho_r(t)}{\delta\langle\zeta_n\rangle^t}
{\rm Tr}\left\{\left(\dot\zeta_n-\varkappa\zeta_n\right)\rho(t)\right\}\nonumber \\
=-\sum_n\frac{\delta\rho_r(t)}{\delta\langle\zeta_n\rangle^t}
{\rm Tr}\left\{\zeta_n\frac{\partial\rho(t)}{\partial t}\right\}
=-\sum_n\frac{\delta\rho_r(t)}{\delta\langle\zeta_n\rangle^t}
{\rm Tr}\left\{\zeta_n\left(i{\hat L}+\varkappa\right)\rho(t)\right\}
\label{20}.
\end{eqnarray}
We define the Kawasaki-Gunton operator \cite{19} in the standard manner
\begin{equation}
{\cal P}_{KG}A = \rho_r(t){\rm Tr}A+
\sum_n\frac{\delta\rho_r(t)}{\delta\langle\zeta_n\rangle^t}
\left\{
{\rm Tr}(A\zeta_n)-\langle\zeta_n\rangle^t{\rm Tr}A
\right\}
\label{21}.
\end{equation}
Using the relations $
{\rm Tr}\{i{\hat L}\rho(t)\}=0$,
${\rm Tr}\left\{{\dot\rho(t)}\right\}
=-\langle\varkappa\rangle^t$, one can prove that the projection operator (21) conserves
its features in the non-Hamiltonian case, namely ${\cal P}_{KG}(t)\rho(t)=\rho_r(t)$,
${\cal P}_{KG}(t){\dot\rho(t)}=
{\dot\rho_r(t)}$. Equation (\ref{20}) can now be rewritten in the projected form
\begin{equation}
\frac{\partial\rho_r(t)}{\partial t}=
-{\cal P}_{KG}(t)\left(i{\hat L}+\varkappa\right)\rho(t)
\label{22}.
\end{equation}
Inserting equation (\ref{22}) into Eq. (\ref{19}) and integrating over time, we obtain
a solution for $\Delta\rho(t)$ as follows

\begin{equation}
\Delta\rho(t)=-\int\limits_{-\infty}^t e^{-\epsilon(t-t')}
U_r^{(\varkappa)}(t,t'){\cal Q}_{KG}(t')\left(
i{\hat L}+\varkappa\right)\rho_r(t')dt'
\label{23},
\end{equation}
where ${\cal Q}_{KG}=1-{\cal P}_{KG}$, and (compare with Eq. (\ref{9}))
\begin{equation}
U_{r}^{(\varkappa)}(t,t')=\exp\left\{
-\int\limits_{t'}^t{\cal Q}_{KG}(t)\left(i{\hat L}+\varkappa\right)
d\tau \right\}
\label{24}.
\end{equation}
Equation (\ref{23}) allows us to
express the right-hand side of Eq. (\ref{11}) in terms of relevant distribution,
i.e. in terms of variables $\langle\zeta_n\rangle^t$.
So we get the closed set of generalized transport equations for observables.
\par
To obtain more transparent equations we define the entropy operator
$$
{\hat S}(t)=\Phi(t)+\sum_n\varphi_n(t)\zeta_n(t) ,
$$
so that $\rho_r(t)=\exp\{-{\hat S}(t)\}$.
Introducing the Mori projection operator \cite{20}

\begin{equation}
{\cal P}_M(t)A=\langle A\rangle_r^t+\sum_n
\frac{\delta\langle A\rangle_r^t}{\delta \langle\zeta_n\rangle^t}
\left(\zeta_n-\langle\zeta_n\rangle^t\right)
\label{26},
\end{equation}
which satisfy the identities ${\cal P}_M^2(t)={\cal P}_M(t)$,
${\cal P}_M(t)\zeta_n=\zeta_n$, one can prove the relation

\begin{equation}
{\cal Q}_{KG}(t)\left\{(i{\hat L}+\varkappa)\rho_r\right\}
={\cal Q}_M(t)\left\{\varkappa-i{\hat L}{\hat S}\right\}\rho_r
\label{27},
\end{equation}
where ${\cal Q}_M(t)=1-{\cal P}_M(t)$ is complementary
to the Mori projector (\ref{26}).
Now the generalized transport equation (\ref{11}) reads
\begin{equation}
\frac{\partial}{\partial t}\langle\zeta_n\rangle^t=
\frac{\partial}{\partial t}\langle\zeta_n\rangle_r^t+{\rm Tr}
\left\{{\cal Q}_M(t)\left(\dot\zeta_n-\varkappa\zeta_n\right)\Delta\rho\right\}
\label{28},
\end{equation}
where $\Delta \rho$ must be taken from equation (\ref{23}). It is useful to
define new dynamical variables (generalized fluxes)
$$
{\cal I}_n(t)={\cal Q}_M(t)\left(\dot\zeta_n-\varkappa\zeta_n\right) ,
$$
which are orthogonal to the space of relevant variables in a sense that ${\cal P}_M{\cal I}_n(t)=0$.
The variables ${\cal I}_n(t)$ determine the effects of macroscopic degrees of freedom of
the evolution on the nonequilibrium state of the non-Hamiltonian system described by a set
of relevant variables and a {\it given} phase space compressibility $\varkappa$.
\par
Incorporating equations (\ref{23}), (\ref{27}), and (\ref{28}) we obtain the final result
for the generalized transport equations
\begin{equation}
\frac{\partial}{\partial t}\langle\zeta_n\rangle^t=
\frac{\partial}{\partial t}\langle\zeta_n\rangle_r^t
+\sum_n\int\limits_{-\infty}^te^{-\epsilon(t-t')}{\cal D}_{mn}(t,t')\varphi_n(t')dt'
+\int\limits_{-\infty}^te^{-\epsilon(t-t')}{\cal F}_{mn}(t,t')dt'
\label{30},
\end{equation}
where

\begin{equation}
{\cal D}_{mn}(t,t')={\rm Tr}\left\{{\cal I}_m(t)U_r^{(\varkappa)}(t,t')
{\cal I}_m(t')\rho_r(t')\right\}
\label{31}
\end{equation}
are the {\it normal} generalized kinetic coefficients, and
\begin{equation}
{\cal F}_{mn}(t,t')={\rm Tr}\left\{{\cal I}_m(t)U_r^{(\varkappa)}(t,t')
{\cal Q}_M(t')\varkappa\left(\sum_n\varphi_n\zeta_n -1\right)\rho_r(t')\right\}
\label{32}
\end{equation}
are the effective forces determining the anomalous contribution into the system
evolution associated with nonlinear phase space compressibility.

\par
The last term in Eq. (\ref{30}) is associated with the initial
irreversibility of the process due to non-Hamiltonian behavior,
while the second term is due to the nonequilibrium process itself.
It does not vanish even when $\varkappa =0$, and coincides with
the well-known result for Hamiltonian ensembles. These terms
demonstrate the general structure of the solution and clearly show
the contributions from two physically different processes. In
practice equations (\ref{31}) and (\ref{32}) can be used with
standard simplifications for particular systems.

\par
In conclusion, we have proposed a possible realization of the
nonequilibrium statistical mechanics for ensembles of particles
obeying non-Hamiltonian dynamics. Starting with the
non-Hamiltonian Poisson bracket (\ref{1}) and the equations of
motion in standard form, we derived the basic relations for
time-dependent dynamical variables with conserved Hamiltonian.
Introducing the reduced description method in analogy with
Hamiltonian dynamics we found the retarded solution (\ref{18}) of
the generalized Liouville equation (\ref{3}). We emphasize
however, that Eq. (\ref{18}) is the specific solution, not
general, and it describes the evolution from a specific auxiliary
relevant distribution (this is reflected in the introduction of
parameter $\epsilon$). Using the projection operator technique we
obtained the generalized transport equations (\ref{30}), which
contain two physically different contributions due to the
statistical irreversibility (\ref{31}) and the
time-irreversibility (\ref{32}) associated with non-Hamiltonian
features of the system.

\acknowledgments{The research is supported by the NSF career award and the ACS petroleum research fund.}


\begin{thebibliography}{24}
\expandafter\ifx\csname
natexlab\endcsname\relax\def\natexlab#1{#1}\fi
\expandafter\ifx\csname bibnamefont\endcsname\relax
  \def\bibnamefont#1{#1}\fi
\expandafter\ifx\csname bibfnamefont\endcsname\relax
  \def\bibfnamefont#1{#1}\fi
\expandafter\ifx\csname citenamefont\endcsname\relax
  \def\citenamefont#1{#1}\fi
\expandafter\ifx\csname url\endcsname\relax
  \def\url#1{\texttt{#1}}\fi
\expandafter\ifx\csname
urlprefix\endcsname\relax\def\urlprefix{URL }\fi
\providecommand{\bibinfo}[2]{#2}
\providecommand{\eprint}[2][]{\url{#2}}

\bibitem[{\citenamefont{Ashurst and Hoover}(1973)}]{1}
\bibinfo{author}{\bibfnamefont{W.~T.} \bibnamefont{Ashurst}} \bibnamefont{and}
  \bibinfo{author}{\bibfnamefont{W.~J.} \bibnamefont{Hoover}},
  \bibinfo{journal}{Phys.\ Rev.\ Lett.} \textbf{\bibinfo{volume}{31}},
  \bibinfo{pages}{206} (\bibinfo{year}{1973}).

\bibitem[{\citenamefont{Andersen}(1980)}]{2}
\bibinfo{author}{\bibfnamefont{H.~C.} \bibnamefont{Andersen}},
  \bibinfo{journal}{J.\ Chem.\ Phys.} \textbf{\bibinfo{volume}{72}},
  \bibinfo{pages}{2384} (\bibinfo{year}{1980}).

\bibitem[{\citenamefont{Evans}(1983)}]{3}
\bibinfo{author}{\bibfnamefont{D.~J.} \bibnamefont{Evans}},
  \bibinfo{journal}{J.\ Chem.\ Phys.} \textbf{\bibinfo{volume}{78}},
  \bibinfo{pages}{3297} (\bibinfo{year}{1983}).

\bibitem[{\citenamefont{Nos\'e}(1984)}]{4}
\bibinfo{author}{\bibfnamefont{S.}~\bibnamefont{Nos\'e}},
  \bibinfo{journal}{Mol.\ Phys.} \textbf{\bibinfo{volume}{52}},
  \bibinfo{pages}{255} (\bibinfo{year}{1984}).

\bibitem[{\citenamefont{Hoover}(1985)}]{5}
\bibinfo{author}{\bibfnamefont{W.~G.} \bibnamefont{Hoover}},
  \bibinfo{journal}{Phys.\ Rev.\ A} \textbf{\bibinfo{volume}{31}},
  \bibinfo{pages}{1695} (\bibinfo{year}{1985}).

\bibitem[{\citenamefont{Martyna et~al.}(1992)\citenamefont{Martyna, Klein, and
  Tuckerman}}]{6}
\bibinfo{author}{\bibfnamefont{G.~J.} \bibnamefont{Martyna}},
  \bibinfo{author}{\bibfnamefont{M.~L.} \bibnamefont{Klein}}, \bibnamefont{and}
  \bibinfo{author}{\bibfnamefont{M.}~\bibnamefont{Tuckerman}},
  \bibinfo{journal}{J.\ Chem.\ Phys.} \textbf{\bibinfo{volume}{97}},
  \bibinfo{pages}{2635} (\bibinfo{year}{1992}).

\bibitem[{\citenamefont{Cao and Voth}(1996)}]{cao1}
\bibinfo{author}{\bibfnamefont{J.~S.} \bibnamefont{Cao}} \bibnamefont{and}
  \bibinfo{author}{\bibfnamefont{G.~A.} \bibnamefont{Voth}},
  \bibinfo{journal}{J.\ Chem.\ Phys.} \textbf{\bibinfo{volume}{105}},
  \bibinfo{pages}{6856} (\bibinfo{year}{1996}).

\bibitem[{\citenamefont{Yang and Cao}(2002)}]{cao2}
\bibinfo{author}{\bibfnamefont{S.~L.} \bibnamefont{Yang}} \bibnamefont{and}
  \bibinfo{author}{\bibfnamefont{J.~S.} \bibnamefont{Cao}},
  \bibinfo{journal}{J.\ Chem.\ Phys.} \textbf{\bibinfo{volume}{117}},
  \bibinfo{pages}{10996} (\bibinfo{year}{2002}).

\bibitem[{\citenamefont{Tuckerman et~al.}(2001)\citenamefont{Tuckerman, Liu,
  Ciccotti, and Martyna}}]{8}
\bibinfo{author}{\bibfnamefont{M.~E.} \bibnamefont{Tuckerman}},
  \bibinfo{author}{\bibfnamefont{Y.}~\bibnamefont{Liu}},
  \bibinfo{author}{\bibfnamefont{G.}~\bibnamefont{Ciccotti}}, \bibnamefont{and}
  \bibinfo{author}{\bibfnamefont{G.~J.} \bibnamefont{Martyna}},
  \bibinfo{journal}{J.\ Chem.\ Phys.} \textbf{\bibinfo{volume}{115}},
  \bibinfo{pages}{1678} (\bibinfo{year}{2001}).

\bibitem[{\citenamefont{Tuckerman et~al.}(1999)\citenamefont{Tuckerman, Mundy,
  and Martyna}}]{9}
\bibinfo{author}{\bibfnamefont{M.~E.} \bibnamefont{Tuckerman}},
  \bibinfo{author}{\bibfnamefont{C.~J.} \bibnamefont{Mundy}}, \bibnamefont{and}
  \bibinfo{author}{\bibfnamefont{G.~J.} \bibnamefont{Martyna}},
  \bibinfo{journal}{Europhys.\ Lett.} \textbf{\bibinfo{volume}{45}},
  \bibinfo{pages}{149} (\bibinfo{year}{1999}).

\bibitem[{\citenamefont{Tully}(1998)}]{tully1}
\bibinfo{author}{\bibfnamefont{J.~C.} \bibnamefont{Tully}},
  \bibinfo{journal}{Faraday Discussions} \textbf{\bibinfo{volume}{110}},
  \bibinfo{pages}{407} (\bibinfo{year}{1998}).

\bibitem[{\citenamefont{Burant and Tully}(2000)}]{tully2}
\bibinfo{author}{\bibfnamefont{J.~C.} \bibnamefont{Burant}} \bibnamefont{and}
  \bibinfo{author}{\bibfnamefont{J.~C.} \bibnamefont{Tully}},
  \bibinfo{journal}{J.\ Chem.\ Phys.} \textbf{\bibinfo{volume}{112}},
  \bibinfo{pages}{6097} (\bibinfo{year}{2000}).

\bibitem[{\citenamefont{Kapral and Ciccotti}(1999)}]{10}
\bibinfo{author}{\bibfnamefont{R.}~\bibnamefont{Kapral}} \bibnamefont{and}
  \bibinfo{author}{\bibfnamefont{G.}~\bibnamefont{Ciccotti}},
  \bibinfo{journal}{J.\ Chem.\ Phys.} \textbf{\bibinfo{volume}{110}},
  \bibinfo{pages}{8919} (\bibinfo{year}{1999}).

\bibitem[{\citenamefont{Nielsen
  et~al.}(2000{\natexlab{a}})\citenamefont{Nielsen, Kapral, and Ciccotti}}]{11}
\bibinfo{author}{\bibfnamefont{S.}~\bibnamefont{Nielsen}},
  \bibinfo{author}{\bibfnamefont{R.}~\bibnamefont{Kapral}}, \bibnamefont{and}
  \bibinfo{author}{\bibfnamefont{G.}~\bibnamefont{Ciccotti}},
  \bibinfo{journal}{J.\ Chem.\ Phys.} \textbf{\bibinfo{volume}{112}},
  \bibinfo{pages}{6543} (\bibinfo{year}{2000}{\natexlab{a}}).

\bibitem[{\citenamefont{Nielsen
  et~al.}(2000{\natexlab{b}})\citenamefont{Nielsen, Kapral, and Ciccotti}}]{12}
\bibinfo{author}{\bibfnamefont{S.}~\bibnamefont{Nielsen}},
  \bibinfo{author}{\bibfnamefont{R.}~\bibnamefont{Kapral}}, \bibnamefont{and}
  \bibinfo{author}{\bibfnamefont{G.}~\bibnamefont{Ciccotti}},
  \bibinfo{journal}{J.\ Chem.\ Phys.} \textbf{\bibinfo{volume}{115}},
  \bibinfo{pages}{5805} (\bibinfo{year}{2000}{\natexlab{b}}).

\bibitem[{\citenamefont{Sergi and Ferrario}(2001)}]{13}
\bibinfo{author}{\bibfnamefont{A.}~\bibnamefont{Sergi}} \bibnamefont{and}
  \bibinfo{author}{\bibfnamefont{M.}~\bibnamefont{Ferrario}},
  \bibinfo{journal}{Phys.\ Rev.\ E} \textbf{\bibinfo{volume}{64}},
  \bibinfo{pages}{056125} (\bibinfo{year}{2001}).

\bibitem[{\citenamefont{Sergi}(2003)}]{14}
\bibinfo{author}{\bibfnamefont{A.}~\bibnamefont{Sergi}},
  \bibinfo{journal}{Phys.\ Rev.\ E} \textbf{\bibinfo{volume}{67}},
  \bibinfo{pages}{021101} (\bibinfo{year}{2003}).

\bibitem[{\citenamefont{Sergi}(2004)}]{ser}
\bibinfo{author}{\bibfnamefont{A.}~\bibnamefont{Sergi}},
  \bibinfo{journal}{Phys.\ Rev.\ E} \textbf{\bibinfo{volume}{69}},
  \bibinfo{pages}{021109} (\bibinfo{year}{2004}).

\bibitem[{\citenamefont{Morrison}(1998)}]{15}
\bibinfo{author}{\bibfnamefont{P.~J.} \bibnamefont{Morrison}},
  \bibinfo{journal}{Rev.\ Mod.\ Phys.} \textbf{\bibinfo{volume}{70}},
  \bibinfo{pages}{467} (\bibinfo{year}{1998}).

\bibitem[{\citenamefont{Zubarev}(1971)}]{16}
\bibinfo{author}{\bibfnamefont{D.}~\bibnamefont{Zubarev}},
  \emph{\bibinfo{title}{Nonequilibrium Statistical Thermodynamics}}
  (\bibinfo{publisher}{Nauka, Moscow}, \bibinfo{year}{1971}).

\bibitem[{\citenamefont{Zubarev et~al.}(1996)\citenamefont{Zubarev, Morozov,
  and R{\"o}pke}}]{17}
\bibinfo{author}{\bibfnamefont{D.}~\bibnamefont{Zubarev}},
  \bibinfo{author}{\bibfnamefont{V.}~\bibnamefont{Morozov}}, \bibnamefont{and}
  \bibinfo{author}{\bibfnamefont{G.}~\bibnamefont{R{\"o}pke}},
  \emph{\bibinfo{title}{Statistical Mechanics of Nonequilibrium Processes}}
  (\bibinfo{publisher}{Akademie Verlag, Berlin}, \bibinfo{year}{1996}).

\bibitem[{\citenamefont{Zubarev}(1996)}]{18}
\bibinfo{author}{\bibfnamefont{D.}~\bibnamefont{Zubarev}},
  \emph{\bibinfo{title}{Modern Methods of Statistical Theory of Nonequilibrium
  Processes. in Science and Technology Report. Modern Problems of Mathematics}}
  (\bibinfo{publisher}{VINITI, Moscow}, \bibinfo{year}{1996}).

\bibitem[{\citenamefont{Kawasaki and Gunton}(1973)}]{19}
\bibinfo{author}{\bibfnamefont{K.}~\bibnamefont{Kawasaki}} \bibnamefont{and}
  \bibinfo{author}{\bibfnamefont{J.~D.} \bibnamefont{Gunton}},
  \bibinfo{journal}{Phys.\ Rev.\ A} \textbf{\bibinfo{volume}{8}},
  \bibinfo{pages}{2048} (\bibinfo{year}{1973}).

\bibitem[{\citenamefont{Mori}(1965)}]{20}
\bibinfo{author}{\bibfnamefont{H.}~\bibnamefont{Mori}},
  \bibinfo{journal}{Prog.\ Theor.\ Phys.} \textbf{\bibinfo{volume}{33}},
  \bibinfo{pages}{423} (\bibinfo{year}{1965}).

\end{thebibliography}
\end{document}